\documentclass[twocolumn,prb,showpacs]{revtex4}
\usepackage{epsfig}
\usepackage{graphics}
\usepackage{graphicx}
\usepackage{dcolumn}
\usepackage{bm}

\begin{document}

\title{Evolution of Superconductivity in Electron-Doped Cuprates:
Magneto-Raman Spectroscopy}
\author{M. M. Qazilbash$^{1,2,\dag}$, A. Koitzsch$^{1,\ddag}$, B. S.
Dennis$^1$, A. Gozar$^{1,\S}$,
Hamza Balci$^2$, C. A. Kendziora$^3$, R. L. Greene$^2$, and G.
Blumberg$^{1,*}$}

\affiliation{
$^{1}$Bell Laboratories, Lucent Technologies, Murray Hill, NJ 07974 \\
$^{2}$Center for Superconductivity Research, Department of Physics,
University of Maryland, College Park, MD 20740\\
$^{3}$United States Naval Research Laboratory, Code 6365, Washington
D.C. 20375}

\date{\today}

\begin{abstract}

The electron-doped cuprates Pr$_{2-x}$Ce$_x$CuO$_{4-\delta}$ (PCCO)
and Nd$_{2-x}$Ce$_x$CuO$_{4-\delta}$ (NCCO) have been studied by
electronic Raman spectroscopy across the entire region of the
superconducting (SC) phase diagram. The SC pairing strength is found
to be consistent with a weak-coupling regime except in the
under-doped region where we observe an in-gap collective mode at
$4.5 k_{B}T_{c}$ while the maximum amplitude of the SC gap is
$\approx 8 k_{B}T_{c}$. In the normal state, doped carriers divide
into coherent quasi-particles (QPs) and carriers that remain
incoherent. The coherent QPs mainly reside in the vicinity of ($\pm
\pi/2a$, $\pm \pi/2a$) regions of the Brillouin zone (BZ). We find
that only coherent QPs contribute to the superfluid density in the
$B_{2g}$ channel. The persistence of SC coherence peaks in the
$B_{2g}$ channel for all dopings implies that superconductivity is
mainly governed by interactions between the hole-like coherent QPs
in the vicinity of ($\pm \pi/2a$, $\pm \pi/2a$) regions of the BZ.
We establish that superconductivity in the electron-doped cuprates
occurs primarily due to pairing and condensation of hole-like
carriers. We have also studied the excitations across the SC gap by
Raman spectroscopy as a function of temperature ($T$) and magnetic
field ($H$) for several different cerium dopings ($x$). Effective
upper critical field lines $H^{*}_{c2}(T, x)$ at which the
superfluid stiffness vanishes and $H^{2\Delta}_{c2}(T, x)$ at which
the SC gap amplitude is suppressed by field have been determined;
$H^{2\Delta}_{c2}(T, x)$ is larger than $H^{*}_{c2}(T, x)$ for all
doping concentrations. The difference between the two quantities
suggests the presence of phase fluctuations that increase for $x
\lesssim 0.15$. It is found that the magnetic field suppresses the
magnitude of the SC gap linearly at surprisingly small fields.

\end{abstract}


\maketitle

\section{Introduction}

The electron doped (\emph{n}-doped) superconducting (SC) cuprates
are an important component in the puzzle of high $T_c$
superconductivity. There is evidence from transport measurements
for both electron-like and hole-like carriers in the
\emph{n}-doped cuprates Pr$_{2-x}$Ce$_x$CuO$_{4-\delta}$ (PCCO)
and Nd$_{2-x}$Ce$_x$CuO$_{4-\delta}$ (NCCO) for Ce dopings in the
vicinity of $x =
0.15$.\cite{ongtwocarrier,wujiang,patrickprl,patrickprb97,gollnik,yoram}
Subsequently, angle-resolved photoemission spectroscopy (ARPES) data
(Fig.~1a-c) show that well-defined electron-like Fermi surface (FS)
pockets exist near the
($\pm \pi/4a$, $\pm \pi/a$) and ($\pm \pi/a$, $\pm \pi/4a$) regions
of the Brillouin zone (BZ) for Ce dopings of $0.10 \leq x \leq
0.15$.
For $x = 0.13$, ARPES intensity map reveals emergence of a hole-like
FS around ($\pm \pi/2a$, $\pm \pi/2a$) regions of
the BZ which spectral weight increases for $x =
0.15$.\cite{Armitage:2002,NCCOARPES}
The Ce doping at which the onset of the superconductivity in the
under-doped side of the phase diagram is observed approximately
coincides with the appearance of the hole-like FS seen by ARPES.
However, the question of whether one
or both, electron- and hole-like carriers are responsible for the
superconductivity has remained unresolved.
Despite the fact that conduction and
superconductivity occur in the copper-oxygen planes in both the
\emph{n}-doped and hole-doped (\emph{p}-doped) cuprates, there are
lingering differences in properties between the two types of
cuprates that require further investigation if one is to arrive at
a comprehensive and unified understanding of the electronic
properties of these materials.

While the magnitude and symmetry of the SC order parameter (or SC
gap) have been thoroughly studied and understood in the
\emph{p}-doped cuprates,\cite{Harlingen,kirtleyreview,kirtley}
similar studies have yet to reach a consensus in the
optimally-doped \emph{n}-doped counterparts. Moreover, there
is disagreement among the experiments that have studied the doping
dependence of the SC order parameter and superfluid density in the
\emph{n}-doped
cuprates.\cite{Amlan,skinta1,skinta2,prozorov2,josephjunc} As far
as the distribution of doped carriers is concerned, ARPES at
optimal doping above $T_c$ indicates the presence of defined
quasi-particles (QPs) and ill-defined incoherent
background.\cite{Armitage:2002,claesson} However, there is a lack
of clear understanding of the relationship between the coherence
properties of introduced carriers and development of the SC order
parameter with doping.

Enhanced temperatures and magnetic fields are known to be
detrimental to superconductivity. The effect of temperature on the
SC order parameter and superfluid density in the high-$T_c$
cuprate superconductors is extensively documented. However, the
influence of magnetic field on these SC properties at temperatures
well below $T_c$ has rarely been investigated spectroscopically,
especially for $n$-doped cuprates. Most studies on the $n$-doped
cuprates have concentrated on the effect of magnetic field on
transport
properties.\cite{ongtwocarrier,wujiang,patrickprl,patrickprb97,gollnik,yoram,Maple,Ong,taillefer,hamzanernst,patrick}
On the theoretical side, researchers have mainly focused on
explaining the effects of temperature on SC
properties.\cite{Hirschfeld,Maki}
Spectroscopic experiments, as well as theoretical frameworks are
necessary for understanding fundamental
properties of high-$T_c$ superconductors in magnetic fields.
These properties are also critical to SC applications.

Here we report a systematic low energy electronic Raman spectroscopy
study of Pr$_{2-x}$Ce$_x$CuO$_{4-\delta}$ (PCCO) and
Nd$_{2-x}$Ce$_x$CuO$_{4-\delta}$ (NCCO) single crystals and films
with different cerium dopings covering the entire SC region of the
phase diagram and determine the magnitude of the order parameter as
a function of doping. We find that the $n$-doped cuprates are in the
$d$-wave weak-coupling regime for Ce dopings greater than or equal
to optimal doping. The under-doped sample is in the strong-coupling
regime and we observe an in-gap collective mode due to strong final
state interactions. We establish that the pairing of coherent
hole-like carriers near ($\pm \pi/2a$, $\pm \pi/2a$) regions of the
BZ leads to superconductivity in the $n$-doped cuprates. We also
compare the coherent part of the ``Raman conductivities"
(Refs.\cite{Sriram,BlumbergRamanDrude}) above and below $T_c$. We
show that a weighted superfluid density can be extracted from the SC
coherence peaks normalized to the frequency shifts. We find that
only coherent QPs contribute to the superfluid density. Moreover, we
study the influence of magnetic field and temperature on
quasi-particle excitations across the SC gap directly by a
spectroscopic method. We plot the variation of the SC gap and
weighted superfluid density as a function of field. We also extract
upper critical field lines $H^{*}_{c2}(T, x)$ at which the
superfluid stiffness vanishes and $H^{2\Delta}_{c2}(T, x)$ at which
the SC amplitude is suppressed by the field. We find a rapid linear
suppression of the SC gap with field.
\begin{figure}[t]
\epsfig{figure=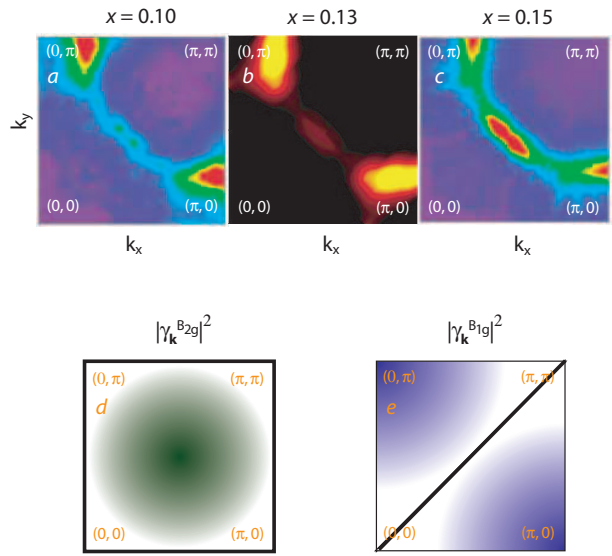,width=80mm} \caption{(color on line) Panels
a-c: Evolution of the Fermi surface of
Nd$_{2-x}$Ce$_x$CuO$_{4-\delta}$ with Ce doping ($x$). Plots of
ARPES intensity near the Fermi energy are reproduced from
Refs.\cite{Armitage:2002} (a and c) and Ref.\cite{NCCOARPES} (b).
Panels (d) and (e) represent momentum dependence of the Raman
coupling $|\gamma_{\textbf{k}}^{(is)}|^2$ in the $B_{2g}$ and
$B_{1g}$ channels respectively. The regions of the Brillouin zone
that contribute to the Raman response in the respective symmetry
channels are shaded. The nodes (regions of zero coupling) in
corresponding symmetry channels are shown in bold.} \label{Fig.1}
\end{figure}

\section{Experimental Methods}

Raman scattering was performed from natural $ab$ surfaces of single
crystals and films of PCCO and single crystals of NCCO. Crystals
with different Ce dopings were grown using a flux
method.\cite{JPeng} After growth, the crystals were annealed in an
Ar-rich atmosphere to induce superconductivity. The SC transitions
were measured by a SQUID magnetometer. The Ce concentration of the
crystals was measured with x-ray wavelength dispersion spectroscopy.
\emph{C}-axis oriented PCCO films were grown on strontium titanate
substrates using pulsed laser deposition.\cite{Maiser,Peng} The
films were grown to a thickness of about 0.8 to 1 $\mu$m to minimize
the substrate contribution to the Raman signal. The SC transitions
were measured by \emph{ac} susceptibility. Rutherford Backscattering
on the films reveals that these thick films are epitaxial and highly
oriented. The films provide an opportunity to study the extremes of
the SC phase because of better control of Ce doping in under-doped
and highly over-doped samples.\cite{Maiser,Peng} 
Numerous previous
studies have established that the phase diagram, crystal structure
and electronic properties of PCCO and NCCO are very similar.

The samples were mounted in an optical continuous helium flow
cryostat. The study of polarization dependence of the Raman
spectra in zero magnetic field was performed in the
pseudo-backscattering geometry with linearly polarized 647 nm and
799 nm excitations from a Kr$^+$ laser. Incident laser powers
between 0.5 and 4~mW were focused to a $50 \times 100$~$\mu$m spot
on the sample surface. For measurements in magnetic field, the
samples were mounted in an optical continuous helium flow cryostat
which was inserted into the bore of a superconducting magnet. The
magnetic field was applied normal to the $\textit{ab}$ plane of
the samples (i.e. along the $c$-axis of the samples). Raman
spectra were measured in a direct backscattering geometry with an
incident wavelength of 647 nm. Incident laser powers between 0.5
and 1~mW were focused to a 50~$\mu$m diameter spot on the sample
surface. The measurements in magnetic field were performed with
circularly polarized light. The spectra displayed in this
manuscript were measured at temperatures between 4 and 30~K by a
custom triple grating spectrometer and the data were corrected for
instrumental spectral response. The sample temperatures quoted in
this work have been corrected for laser heating.

\section{Raman scattering symmetries}

The polarization directions of the incident, \textbf{e}$_i$, and
scattered, \textbf{e}$_s$, photons are indicated by
(\textbf{e}$_i$\textbf{e}$_s$) with $x=[100]$, $y=[010]$,
$x'=[110]$, $y'=[\overline{1}10]$, $R=x+i\,y$ and $L=x-i\,y$. The
data were obtained in $(xy)$, $(x'y')$, $(xx)$, $(RR)$ and $(RL)$
scattering geometries. For the tetragonal $D_{4h}$ symmetry of the
\emph{n}-doped cuprates, these geometries correspond to
$B_{2g}$+$A_{2g}$, $B_{1g}$+$A_{2g}$ $A_{1g}$+$B_{1g}$,
$A_{1g}$+$A_{2g}$ and $B_{1g}$+$B_{2g}$ representations. Using
circularly polarized light we confirmed that the contribution to
the $A_{2g}$ channel is very weak for both PCCO and NCCO. The
spectra in $(x'y')$ scattering geometry were subtracted from the
spectra in the $(xx)$ scattering geometry to obtain the $A_{1g}$
Raman response.

The electronic Raman response function, $\chi''^{(is)}(\omega)$,
for a given polarization geometry (\textbf{e}$_i$\textbf{e}$_s$)
is proportional to the sum over the density of states at the FS
weighted by the square of the momentum (\textbf{k})
dependent Raman vertex
$\gamma_{\textbf{k}}^{(is)}$.\cite{Dierker,Hackl,Klein,Devereaux1}
Because the scattering geometries selectively discriminate between
different regions of the FS, electronic Raman spectroscopy
provides information about both the magnitude and the \textbf{k}
dependence of the SC OP. In the effective mass approximation
$\gamma_{\bf{k}}^{B_{1g}} \propto t(\cos k_{x}a - \cos k_{y}a)$
and $\gamma_{\bf{k}}^{B_{2g}} \propto 4t' (\sin k_{x}a \sin
k_{y}a)$ where $t$ and $t'$ are nearest and next-nearest neighbor
hopping integrals in the tight-binding model.
For the $B_{2g}$
channel, the Raman vertex is most sensitive to ($\pm \pi/2a$, $\pm
\pi/2a$) regions of the BZ and vanishes along (0,
0)$\rightarrow$($\pi/a$, 0) and equivalent lines. For the $B_{1g}$
channel, nodal (0, 0)$\rightarrow$($\pi/a$, $\pi/a$) diagonals do
not contribute to the intensity that mainly integrates from
regions near intersections of the FS and the BZ boundary. The
pictorial representations of the $B_{2g}$ and $B_{1g}$ channels
are shown in Fig. 1d,e.
On comparison with ARPES data, one can see
that the $B_{2g}$ scattering channel probes the hole-like
pockets in the vicinity of the ($\pm \pi/2a$, $\pm \pi/2a$)
regions of the BZ while Raman intensity in the $B_{1g}$ channel
originates from the electron-like FS near the ($\pm \pi/4a$,
$\pm \pi/a$) and ($\pm \pi/a$, $\pm \pi/4a$) regions of the BZ.
The unscreened Raman response in $A_{1g}$ channel does not have
symmetry imposed nodal lines and measures an overall average
throughout the BZ.
However, the full symmetric response is expected to be
screened by Coulomb interaction induced charge backflow that
redistributes and strongly suppresses the spectral intensity.
The screening is expected to be weaker if the Raman vertex
$\gamma_{\textbf{k}}^{A_{1g}}$ is rapidly changing with
wavevector.\cite{Devereaux96}

\section{Doping and Polarization Dependence}

In Fig. 2 we show doping dependence of the low energy electronic
Raman response of PCCO single crystals and films.
One can see that the Raman scattering intensity is
significantly stronger in the $B_{2g}$ than in the $B_{1g}$
channel for all Ce concentrations. This is only partly due to
resonance resulting from inter-band
transitions.\cite{BlumbergNCCO} More importantly, it underlines
the significance of next-nearest neighbor hopping $t'$ in
\emph{n}-doped cuprates and is in contrast with \emph{p}-doped
cuprates where the response in the $B_{2g}$ channel is generally
weaker than in $B_{1g}$.\cite{Liu} Coulomb screening should lead
to a much weaker Raman response in the fully symmetric $A_{1g}$
channel. However, we find that the intensities in the $A_{1g}$
channel are of the same order of magnitude as those in the
non-symmetric $B_{1g}$ and $B_{2g}$ channels in the over-doped samples
and are significantly stronger in the under- and optimally-doped samples.
This lack of screening is possible if the effective mass on the FS
changes sign.\cite{Cardona} This is the case in the electron-doped
cuprates where strong evidence exists for both electron- and
hole-like
carriers.\cite{ongtwocarrier,wujiang,patrickprl,yoram,Armitage:2002,NCCOARPES}
\begin{figure*}[t]
\epsfig{figure=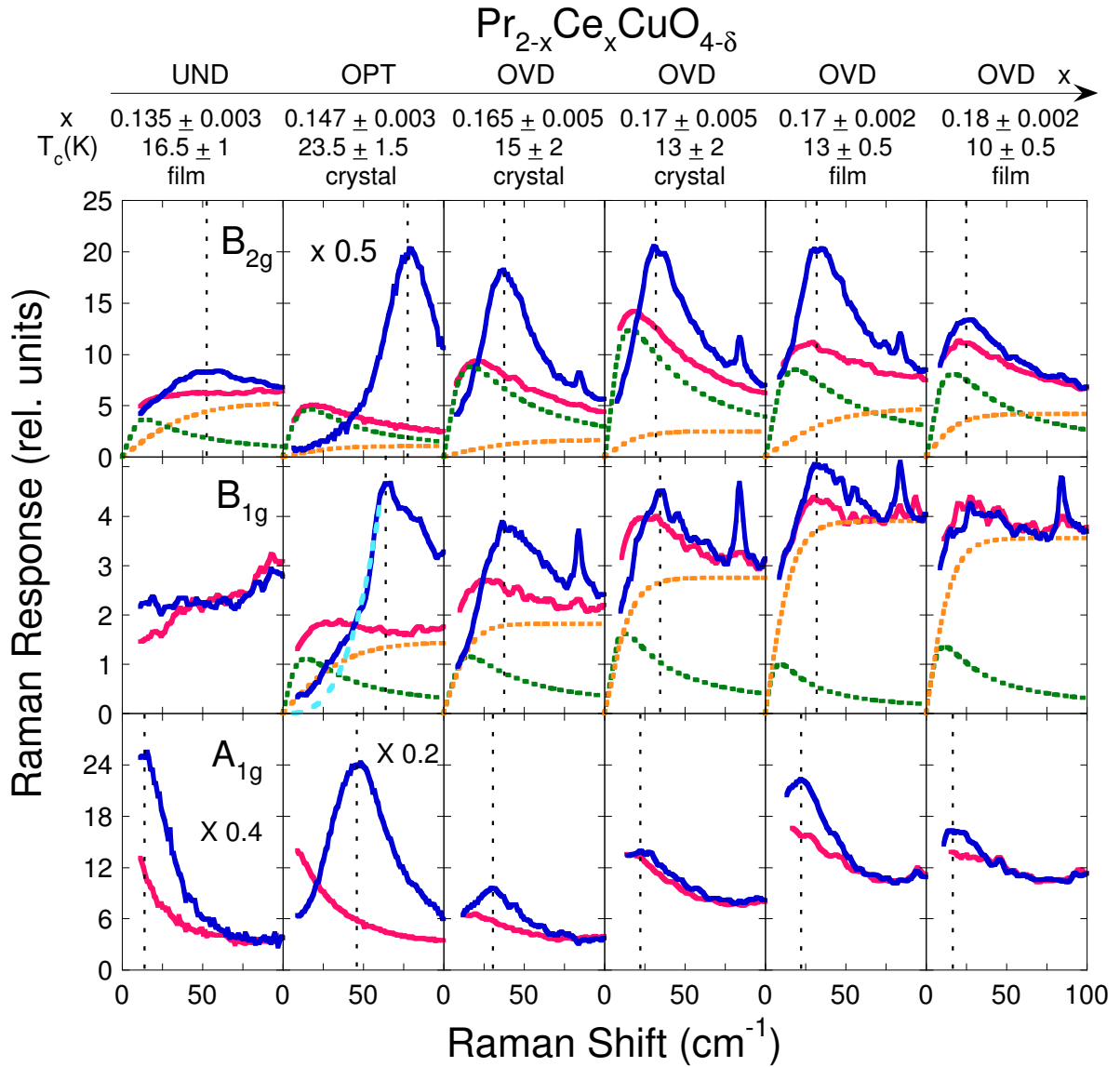,width=160mm} \caption{(color online) Doping
dependence of the low energy electronic Raman response of PCCO
single crystals and films for $B_{2g}$, $B_{1g}$ and $A_{1g}$
channels obtained with 647~nm excitation. The columns are arranged
from left to right in order of increasing cerium doping.
Abbreviations UND, OPT and OVD refer to under-doped, optimally doped
and over-doped samples respectively. The light (red) curves are the
data taken just above the respective $T_c$ of the samples. The
normal state response in the $B_{2g}$ and $B_{1g}$ channels is
decomposed into a coherent Drude contribution (green dotted line)
and an incoherent continuum (yellow dotted line). The dark (blue)
curves show the data taken in the SC state at $T$~$\approx$~4~K. The
dashed vertical lines indicate positions of the SC coherence peaks.
For the optimally-doped crystal a low-frequency $\omega^3$ power law
is shown in the $B_{1g}$ panel for comparison (light-blue dotted
line). Over-doped crystals and films show similar behavior,
indicating the good quality of the films.}\label{Fig.2}
\end{figure*}

We decompose the Raman response in the normal state into two parts, a
featureless continuum and a low-frequency quasi-elastic scattering
peak (QEP):
\begin{equation}
     \chi''_N(\omega) = \chi''_{QEP}(\omega) + \chi''_{MFL}(\omega).
     \label{chiN}
\end{equation}
The QEP response
\begin{equation}
     \chi''_{QEP}(\omega) = a^{(is)} \frac{\Gamma \omega}{\omega^2 + \Gamma^2}
     \label{chiQEP}
\end{equation}
is described in a Drude model as QP contribution from
doped carriers\cite{BlumbergRamanDrude,Andreas} while the
featureless continuum
\begin{equation}
     \chi''_{MFL}(\omega) = b^{(is)} \tanh({\omega}/{\omega_c})
     \label{chiMFL}
\end{equation}
represents a collective incoherent response.\cite{Varma} Symmetry
dependent $a^{(is)}$ and $b^{(is)}$ parameters control the
spectral weight in these coherent and incoherent channels,
$\omega_c$ is a cut-off frequency of order
$k_{B}T$~(Ref.\cite{Varma}) and the QEP position $\Gamma$ is the
Drude scattering rate that at low temperatures is about 2 meV for
the entire doping range studied.
The Drude part describes the relaxation dynamics of the QPs that results
from electron-electron interactions and is therefore strongly
temperature dependent\cite{Andreas} in contrast to scattering from
impurities which is expected to be independent of temperature.
This deconvolution of the Raman
response into two components presented here is consistent with the
ARPES data that displays defined QPs as well as ill-defined
excitations in different parts of the
FS.\cite{Armitage:2002,claesson}
One can observe from the
deconvolution that the Raman response in the $B_{2g}$ channel is
dominated by the QP (Drude) response while the $B_{1g}$ channel is
dominated by the incoherent continuum.\cite{films}
The well-defined hole-like QP states reside mainly in the
vicinity of the ($\pm \pi/2a$, $\pm \pi/2a$) regions of the BZ.
The evolution of the integrated QP spectral
weight of the ``Raman conductivity"
(Refs.\cite{Sriram,BlumbergRamanDrude})
\begin{equation}
     I_{N}^{B_{2g}}(x)=\int (\chi''^{B_{2g}}_{QEP}/{\omega})\,d\omega
     \label{IN}
\end{equation}
as a function of Ce doping $x$ is shown in Fig.~4(c).
While on the under-doped side, $I_{N}^{B_{2g}}(x)$ exhibits the
expected increase
proportional to $x$, the integrated coherent contribution
saturates above optimal doping $x \gtrsim 0.145$.
At higher Ce doping, additional carriers contribute mainly to the
electron-like incoherent
response and can be observed as an increasing intensity of the
featureless Raman continuum $\chi''_{MFL}(\omega)$, particularly
in the $B_{1g}$ channel.

\section{Pair breaking excitations}

In the SC state, the strength of the low-frequency intensity in
the normal state is reduced and the spectral weight moves to the
$2\Delta$ coherence peak resulting from excitations out of the SC
condensate. In the $B_{2g}$ and $A_{1g}$ channels, the
``pair-breaking" SC coherence peaks appear for all dopings while
in the $B_{1g}$ channel these SC coherence peaks are negligibly
weak in the under- and the most over-doped films.
For the optimally-doped crystal (T$_c$
= 23.5 K), the SC coherence peak energy is larger in the $B_{2g}$
channel compared with that in $B_{1g}$, and for all channels it is
larger than the scattering rate $\Gamma$ obtained from the spectra
in the normal state.
The intensity below the SC coherence peaks vanishes smoothly without a
threshold to the lowest frequency measured. The absence of a
threshold that has been observed in $s$-wave superconductors
precludes interpretation in terms of a fully gapped
FS.\cite{Dierker,Hackl} The smooth decrease in the Raman response
below the SC coherence peak is consistent with nodes in the gap.
We compare the low-frequency tail in the $B_{1g}$ response (Fig.
2) to an $\omega^3$ power law that is expected for a
$d_{x^{2}-y^{2}}$-wave superconductor in the clean
limit.\cite{Devereaux2} The observed deviation from a cubic to a
linear response at the lowest frequencies is an indication of
low-energy QP scattering.\cite{Devereaux3} The data for
optimally-doped PCCO
is very similar to that for optimally-doped NCCO (see Fig.~3) which
was interpreted in terms of a non-monotonic $d$-wave order parameter
with nodes along the (0, 0)$\rightarrow$($\pi/a$, $\pi/a$)
diagonal and the maximum gap being closer to this diagonal than to
the BZ boundaries.\cite{BlumbergNCCO} Recent evidence from ARPES
that the SC gap maximum is located closer to the nodal direction
in optimally doped samples confirms the interpretation of the
Raman data.~\cite{Takahashi}

\begin{figure}[t]
\epsfig{figure=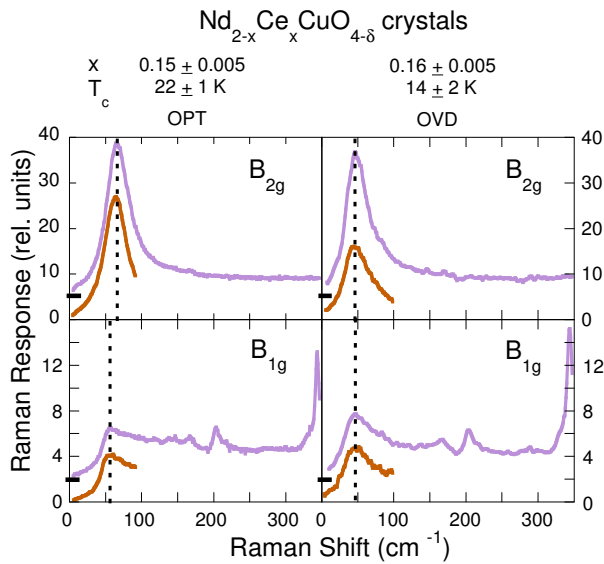,width=80mm}
\caption{(color online) A comparison of the Raman response in the
superconducting state for an optimally doped (OPT) NCCO crystal in
the first column and an over-doped (OVD) NCCO crystal in the second
column. The first and second rows show spectra for $B_{2g}$ and
$B_{1g}$ channels respectively. The violet (light) and orange (dark)
curves are data taken with red laser excitation ($\lambda_L$ = 647
nm) and near-IR excitation ($\lambda_L$ = 799 nm). The $B_{2g}$ and
B$_{1g}$ data for $\lambda_L$ = 647 nm has been shifted up by 5
units and 2 units respectively. All spectra are taken at
$T$~$\approx$ 4~K. The dashed vertical lines show the positions of
the SC coherence peaks. Data with near-IR laser excitation was
obtained up to 100~cm$^{-1}$ Raman shift.} \label{Fig.3}
\end{figure}
Interestingly, in the over-doped PCCO samples (Fig.~2) and over-doped
NCCO crystal (Fig.~3), the SC coherence peak positions are at the same
energies for both the $B_{2g}$ and $B_{1g}$ channels. The peak
positions and intensities decrease in the over-doped regime compared
to the optimally-doped samples.
Moreover, the peak energies are similar to
$\Gamma$ indicating that superconductivity is departing from the
clean limit.\cite{Devereaux3} The Raman response below the SC
coherence peaks vanishes smoothly and no well-defined threshold is
observed. The data for the 799~nm excitation is measured down to
4.5~cm$^{-1}$ and shows no obvious sub-gap threshold. The peak
positions and the sub-gap Raman response in the NCCO crystals are
almost independent of the laser excitation energies (Fig.~3).
Similar symmetry independent ``pair-breaking" peak energies with
continuously decreasing Raman scattering intensity down to low
frequencies have been observed in the Raman spectra in over-doped
samples of \emph{p}-doped Bi-2212.\cite{kendziora,hacklSPIE,tajima}
The Raman data presented in Figs.~2-3 for over- \emph{n}-doped samples
is similar to the Raman data for over-doped Bi-2212.
The coincidence of
coherence peak energies in the $B_{1g}$ and $B_{2g}$ channels may
be caused by enhanced QP scattering,\cite{Devereaux3} although
this remains an open question. Nevertheless, the smooth and
continuous decrease in Raman intensity below the coherence peak is
consistent with a nodal gap structure.\cite{prozorov2,josephjunc}

In Fig. 4(a) we show that the SC coherence peak energy (2$\Delta$)
has a pronounced maximum at optimal doping. For comparison, we
include the value of twice the SC gap energy obtained from point
contact tunneling spectroscopy.\cite{Amlan, Qazilbash}
For optimally- and over-doped samples the maximum values of the Raman
SC coherence peak
energies are very similar to the single particle spectroscopy gap
values and can therefore be associated with twice the SC gap
magnitude. This is not the case for under-doped samples where the
tunneling spectroscopy data exhibits a gap (2$\Delta$) that is
larger than the Raman SC coherence peak energy of 52~cm$^{-1}$
($\approx$ 6.4 meV). The normalized tunneling gap value in Fig. 4b
suggests that the under-doped \emph{n}-doped cuprates are in the strong
coupling regime. Therefore, in under-doped samples the two QPs excited out
of the SC condensate by Raman processes continue to interact,
binding into a collective excitonic state that costs less energy
than excitation of two independent QPs. The SC coherence peak in
the $B_{2g}$ channel in the under-doped sample is actually an in-gap
collective mode. Similar observations were made previously in the
\emph{p}-doped cuprates in the $B_{1g}$
channel.\cite{Liu,Blumberg:97,Blumberg:98} The significance of
final state interactions in the formation of a collective mode in
under-doped $p$-doped cuprates has been demonstrated in Refs.
\cite{BardasisSchrieffer:61,Chubukov:99,Chubukov:05} The important
difference here is that in the $n$-doped cuprates the collective
mode appears in the $B_{2g}$ channel.

We note for the $x = 0.13$ sample the first appearance of the SC
coherence peak in the $B_{2g}$ channel coincides with the appearance
of a hole-like FS near the ($\pm \pi/2a$, $\pm \pi/2a$) regions of
the BZ as viewed by ARPES (see Fig. 1b and Ref.\cite{NCCOARPES})
while the response in the $B_{1g}$ channel does not show any
comparable signatures of superconductivity. These observations
relate superconductivity to the appearance of the hole-like FS. The
electron-like carriers that are clearly present near the ($\pm
\pi/4a$, $\pm \pi/a$) and ($\pm \pi/a$, $\pm \pi/4a$) regions of the
FS do not show contribution to superconducting pairing. When Ce
doping is increased to $x \approx 0.15$, ARPES data shows that the
electron pocket persists and the excitations around hole-like FS
acquires more coherent spectral weight.\cite{Armitage:2002}
Well-defined SC coherence peaks appear in both $B_{2g}$ and $B_{1g}$
channels with the peak energy about twenty percent higher in the
$B_{2g}$ channel. One can infer that for optimal doping the
superconductivity exists in both hole and electron bands but the gap
is larger in the hole channel. For over-doped samples, it is likely
that the FS is hole-like and centered at ($\pm \pi/a$, $\pm \pi/a$)
points. The evolutionary trend of the FS with doping in the ARPES
data supports this view though ARPES data on over-doped samples is
still lacking. The situation may be more complicated because there
is evidence from Hall effect data that both hole-like and
electron-like carriers exist even for over-doped samples with $x =
0.17$.\cite{patrickprl,yoram} Nevertheless, hole-like carriers
dominate the low temperature Hall effect data for dopings $x \geq
0.16$. Therefore, one can hypothesize that in the over-doped
samples, condensation of hole-like carriers is primarily responsible
for the SC coherence peaks in all Raman scattering symmetries.

\begin{figure}[t]
\epsfig{figure=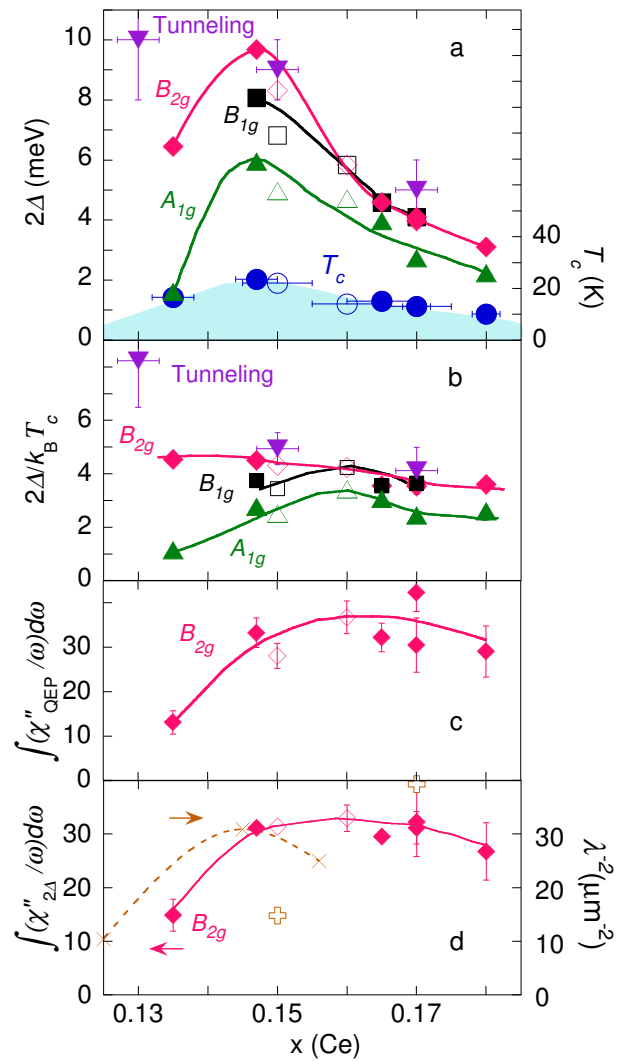,width=80mm} \caption{(color online) The
phase diagram of PCCO (filled symbols) and NCCO (open symbols)
superconductors explored by Raman spectroscopy. Panels show: (a)
T$_c$ and 2$\Delta$ peak positions for $B_{1g}$, $B_{2g}$ and
$A_{1g}$ channels as well as the separation between coherence
peaks from point contact tunneling
spectroscopy;\cite{Amlan,Qazilbash} (b) Magnitude of the
normalized SC coherence peaks (2$\Delta/k_{B}T_{c}$) from three
Raman channels, and from tunneling;\cite{Amlan,Qazilbash} (c) The
integrated quasi-particle (Drude) response just above $T_{c}$; (d)
Integrated intensity of reduced 2$\Delta$ coherence peaks
($\chi''_{2\Delta}/\omega$) at 4~K. For comparison, we plot
1/$\lambda(0)^2$ values from ref.\cite{skinta1} ($\times$) and
ref.\cite{prozorov2} (open crosses). Error bars on the cerium
concentrations are shown only on the T$_c$ data points to preserve
clarity of the figure. Solid lines are guides to the eye.}
\label{Fig.4}
\end{figure}
Reduced energies of the Raman SC coherence peaks,
$2\Delta/k_{B}T_{c}$, are plotted in Fig. 4b as a function of
doping. For the channel that exhibits the highest ratio, $B_{2g}$,
the values fall between 4.5 for the optimally-doped samples and 3.5
for the most over-doped samples.
These ratios are consistent with those inferred
from electron tunneling and infrared reflectivity
measurements,\cite{Amlan,Qazilbash,zimmers} and within the
prediction of the mean-field BCS values for \emph{d}-wave
superconductors.\cite{Maki}. The coherence peak energy remains
below $4.2k_{B}T_{c}$ for the $B_{1g}$ channel and is even lower
for the $A_{1g}$ channel. The reduced energies for all the
channels are lower than for \emph{p}-doped
materials\cite{Liu,Blumberg:97,Blumberg:98,Kendziora:95}
suggesting a $d$-wave BCS weak coupling limit in the
\emph{n}-doped cuprates for optimally- and over-doped samples.

\section{Raman sum-rule}

In Fig.~4(d) we plot the integrated reduced coherence peak
intensity in the SC state,
\begin{equation}
     I_{SC}^{B_{2g}}(x)=\int({\chi''^{B_{2g}}_{2\Delta}}/{\omega})\,d\omega,
     \label{}
\end{equation}
where $\chi''^{B_{2g}}_{2\Delta}(\omega)$ is the SC coherence
response with the incoherent continuum subtracted. The details of
the analysis are given in the appendix. For the non-symmetric
channels in $T \rightarrow 0$ limit
\begin{equation}
     I_{SC}^{(is)} \propto
     \sum_{\bf{k}}{(\gamma_{\bf{k}}^{(is)})^2
     \frac{{\Delta_{\bf{k}}}^2}{{2E_{\bf{k}}}^3}}
     \label{}
\end{equation}
is proportional to the superfluid density
\begin{equation}
     \rho_s \propto
     \sum_{\bf{k}}{\frac{{\Delta_{\bf{k}}}^2}{{2E_{\bf{k}}}^3}}
     \label{}
\end{equation}
weighted by the square of the Raman coupling
vertex.\cite{muzikar,Maki} Here $E_{\bf{k}}$ is the QP dispersion
in the SC state and $\Delta_{\bf{k}}$ is the SC gap. The
superfluid densities ($\rho_s\propto1/\lambda^2$) obtained from
penetration depth ($\lambda$) measurements\cite{skinta1,prozorov2}
are plotted in Fig.~4(d) for comparison\cite{resonance}.
We note
that the values of the integrated reduced coherence intensities in
the $B_{2g}$ channel do
not change from the normal to SC state (Fig.~4c and d)
demonstrating a partial sum rule similar to the
Ferrell-Glover-Tinkham sum-rule in optics. Also, the
$I_{N}^{B_{2g}}(x)=I_{SC}^{B_{2g}}(x)$ equality implies that only
Drude QPs control the superfluid density and that the electron-like
incoherent carriers doped above optimal doping do not contribute to the
superfluid stiffness.

\begin{figure*}[t]
\epsfig{figure=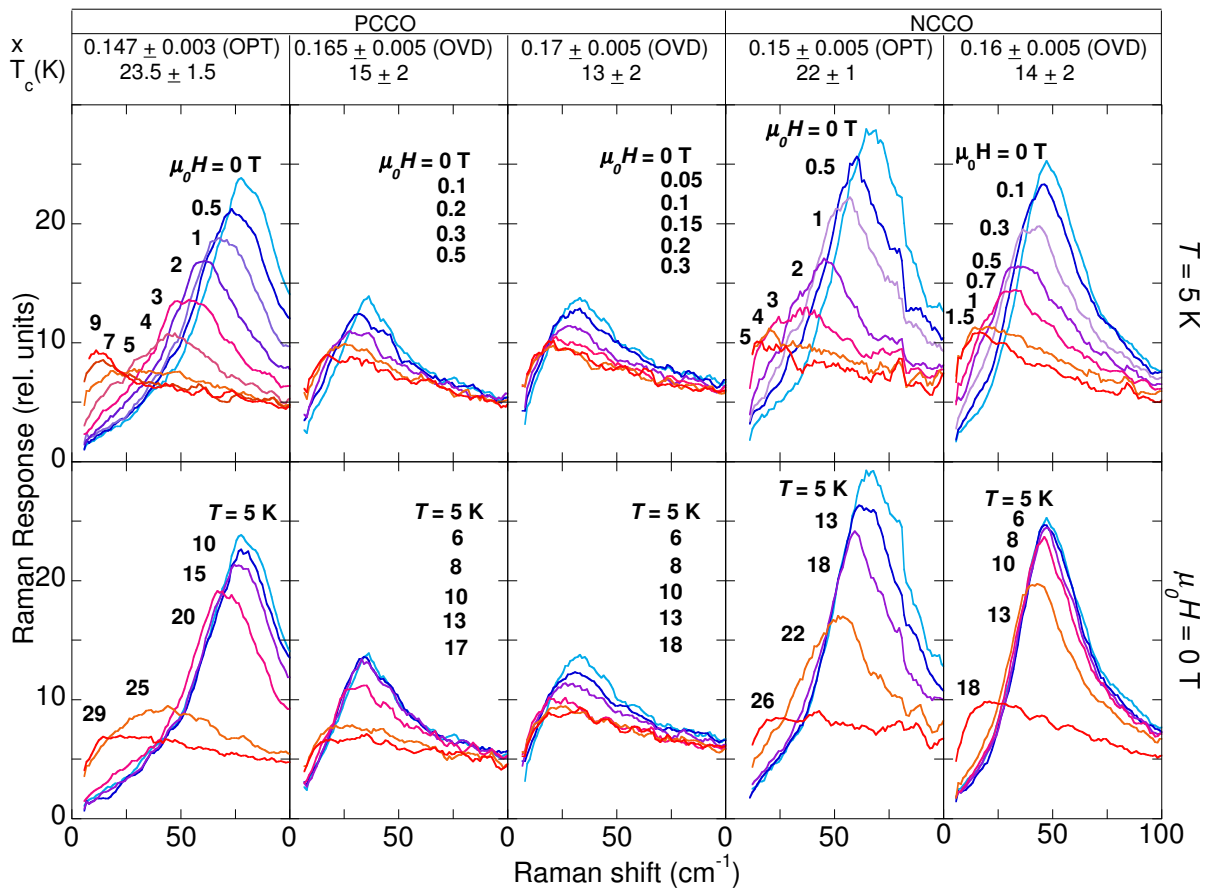,width=160mm} \caption{(color online) Raman
response function $\chi"(\omega,H,T)$ for 647~nm excitation and RL
polarization for five single crystals of PCCO and NCCO with
different Ce dopings $x$. Abbreviations OPT and OVD stand for
optimally doped and over-doped samples respectively. The first row
shows the disappearance of the 2$\Delta$ coherence peak in
increasing magnetic field applied normal to the $ab$-plane of the
crystals at 5~K. The second row shows the temperature dependence of
the 2$\Delta$ peak in zero magnetic field.} \label{Fig.5}
\end{figure*}
Certain aspects of the observations in the preceding paragraph
need further comment.
For conventional superconductors that are Fermi liquids above
$T_c$, theory does not predict a partial sum-rule for low energy
Raman scattering connecting the normal and SC states akin to the
Ferrell-Glover-Tinkham sum-rule in optics. However, we have
experimentally demonstrated the existence of a partial Raman
sum-rule at low energies in the normal and SC states for the
$n$-doped cuprates.
The question of whether this partial sum-rule is limited to the
$n$-doped cuprates or is more generic encompassing $p$-doped
cuprates as well can be decided by a similar analysis of the Raman
data that exists for several families of $p$-doped cuprates.

\section{Effects of temperature and magnetic field}

Circularly polarized light in the right-left (RL) scattering
geometry was used for the data displayed in this section. For the
tetragonal $D_{4h}$ symmetry, this geometry corresponds to the sum
of $B_{1g}$+$B_{2g}$ representations. It has been shown in
previous Raman measurements that for 647 nm incident laser light
the scattering in the $B_{2g}$ channel is resonantly enhanced and
is about an order of magnitude greater than scattering intensity
in the $B_{1g}$ channel.\cite{BlumbergNCCO} Therefore, the
$B_{2g}$ channel dominates the spectra for right-left
polarization.

To recap, the QP excitations across the SC gap lead to a
pair-breaking 2$\Delta$ coherence peak close to twice the gap
energy. The 2$\Delta$ peak is a measure of the magnitude of the SC
order parameter for optimally- and over-doped samples, while the
integrated intensity of the reduced 2$\Delta$ coherence peak
\begin{equation}
     I_{SC}(H, T) = \int({\chi''_{2\Delta}(\omega,H,T)}/{\omega})\,d\omega
     \label{}
\end{equation}
is related to the superfluid density $\rho_s$. This is the
framework that forms a basis for analysis of the Raman data in
this section.

\begin{figure*}[t]
\epsfig{figure=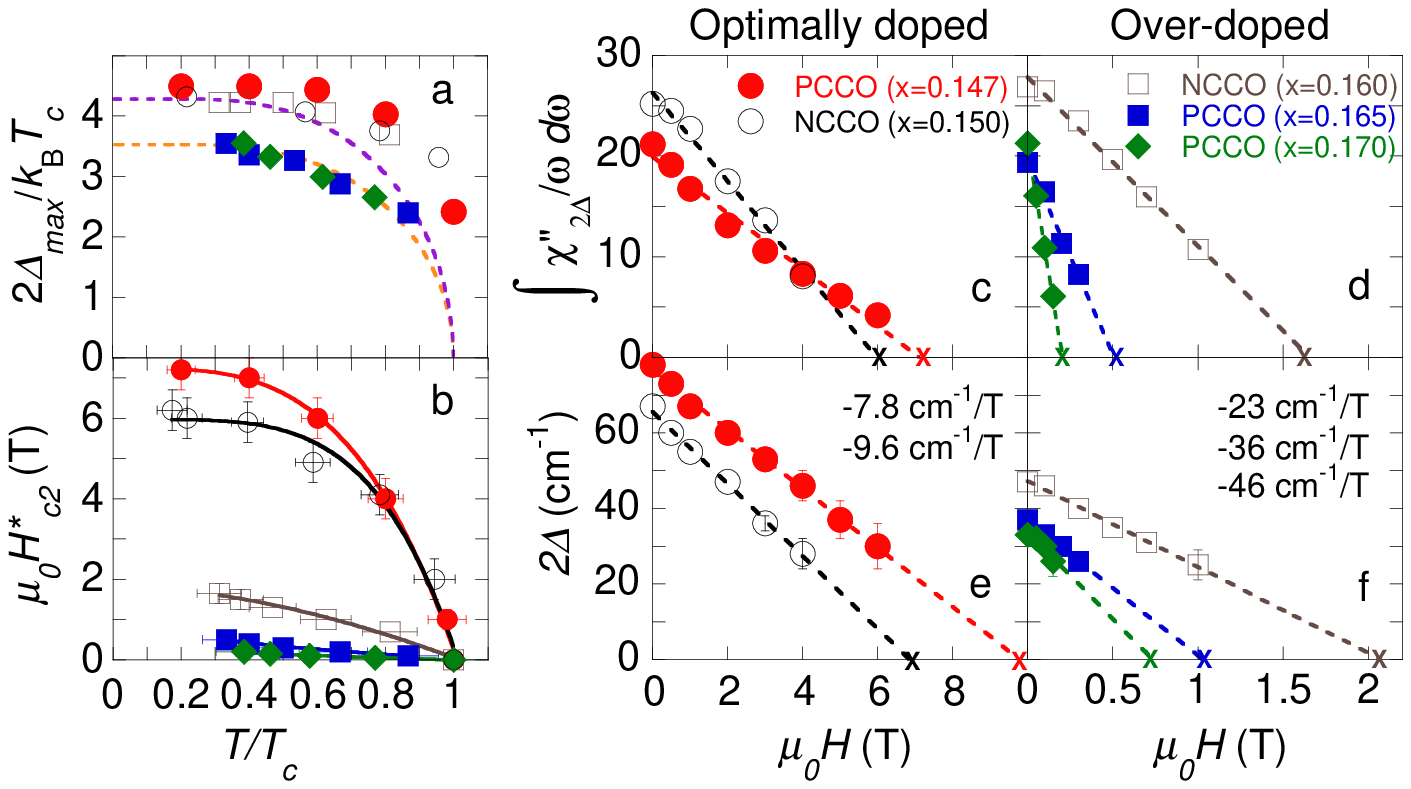,width=130mm} \caption{(color online) The
temperature and field dependence of the SC order parameter
amplitude and stiffness for two NCCO (with Ce dopings $x = 0.15$
and 0.16; open symbols) and three PCCO ($x = 0.147$, 0.165 and
0.17; filled symbols) single crystals. The same symbols are used
for all plots to indicate data on a particular crystal. The
temperature dependence (a) of the reduced SC gap magnitude at zero
field, and (b) of the effective upper critical field $H^{*}_{c2}$
that completely suppresses the SC coherence peak intensity at the
given temperature. In plot (a) the dotted lines are weak-coupling
BCS predictions for temperature dependence of $d$-wave (violet)
and $s$-wave (yellow) SC gaps. Horizontal error bars are not shown
in Fig. 6a and are the same as in Fig. 6b. The field dependence at
5~K of the integrated reduced SC coherence intensity, $I_{SC}(H)$,
is shown in panels (c) and (d) and the SC coherence peak energy
(2$\Delta$) in panels (e) and (f) for optimally doped and
over-doped crystals correspondingly.
``x" indicates the effective upper critical fields ($H^{*}_{c2}$) and
the field values of the SC gap collapse ($H^{2\Delta}_{c2}$)
extrapolated from the data.
Panels (e) and (f) also include the rate of the gap suppression
$\frac{\partial\,{2\Delta(H,T)}}{\partial\,H}$. } \label{Fig.6}
\end{figure*}
Fig.~5 exhibits the field and temperature dependence
of the SC coherence peak at the maximum gap value
(2$\Delta_{max}$) for the optimally-doped ($x \approx 0.15$) and
over-doped ($x > 0.15$) PCCO and NCCO crystals. The coherence peak
loses intensity and moves to lower energies by either increasing
the temperature or magnetic field. We define an effective upper
critical field, $H^{*}_{c2}(T, x)$, as the field that completely
suppresses the coherence peak intensity: the Raman response
remains field independent for $H > H^{*}_{c2}$. Above $T_c$ or
$H^{*}_{c2}$ the SC coherence peak vanishes and the Raman response
acquires a low-frequency QEP. The non-SC response is similar for
both cases: $T
>  T_c$ and $H > H^{*}_{c2}$ at the lowest temperature, and
therefore is independent of the means used to quench
superconductivity. As discussed previously, the Raman response in
the normal state can be described by a QEP Drude response of doped
QP carriers above an incoherent featureless continuum, eqs. (1-3).
At low temperatures $\Gamma(x) \approx 2$~meV for the entire studied
doping range.

Reduced gap values, 2$\Delta_{max}/k_BT_c$, from Raman data in
zero field on five single crystals of various doping levels are
plotted in Fig.~6a as a function of the reduced temperature
($T/T_c$). For the lowest measured temperature,
2$\Delta_{max}/k_BT_c$ values fall between 4.5 for the optimally
doped crystals and 3.5 for the most over-doped PCCO crystals. For
optimally doped samples, the gap appears to open up faster than
the mean-field BCS prediction\cite{Maki} as the temperature is
reduced below $T_c$ suggesting phase fluctuations and a finite
pairing amplitude above $T_c$. The temperature dependence of the
gap energy for the most over-doped samples is close to the
prediction of BCS theory.\cite{d_vs_s}

The mixed state consists of normal regions inside vortex cores
coexisting with SC regions surrounding the vortices. The Raman
response in a field, $\chi''(\omega, H, T)$, is assumed to be a
sum of contributions from the SC and normal regions. The normal
state contribution to the measured Raman spectra is further
assumed to be proportional to the number of vortices which itself
is proportional to the applied field:
\begin{equation}
\chi''(\omega, H, T) =  \chi''_{SC}(\omega, H, T)  +
\frac{H}{H^{*}_{c2}} \chi''_{N}(\omega, T).
     \label{chiHT}
\end{equation}
Here $\chi''_{N}(\omega, T)$
is the normal state Raman response at or
above the critical field value $H^{*}_{c2}(T)$. We extract the
Raman response of the SC regions $\chi''_{SC}(\omega, H, T)$ from
the data using eq.~(\ref{chiHT}).
We then decompose $\chi''_{SC}(\omega, H,
T)$ into a low-frequency SC 2$\Delta$ coherence peak,
$\chi''_{2\Delta}(\omega, H, T)$, and a featureless continuum
above the peak frequency.
$H^{*}_{c2}$ in eq.~(\ref{chiHT}) is the estimate
of the upper critical field at which the SC coherence peak
vanishes in the raw data.

\section{Upper critical fields and superconducting coherence
length}

Estimates of $H^{*}_{c2}$ for different temperatures and Ce
concentrations are plotted vs. $T/T_c$ in Fig.~6b. $H^{*}_{c2}(T)$
displays negative curvature to the lowest measured temperature (5
K). $H^{*}_{c2}(T)$ saturates at low temperatures for samples near
optimal doping. With increasing doping we observe a dramatic
reduction of $H^{*}_{c2}$ in the entire temperature range.

In Fig.~6c,d we plot for optimally- and over-doped crystals, the
integrated coherence intensity $I_{SC}(H, T)$ at $T \approx 5$~K,
as a function of field.
$I_{SC}(H, T)$ is proportional to the
superfluid density $\rho_s$ weighted by the square of the Raman
coupling vertex (eqs. (6)-(7)).
In the panels Fig.~6e,f the coherence peak energy
$2\Delta(H)$ is displayed as a function of field at $T$ $\approx$
5~K. It appears that both the superfluid stiffness and the SC gap
magnitude show a monotonic almost linear decrease with field. We
use a linear continuation to determine critical values of
$H^{*}_{c2}$ and $H^{2\Delta}_{c2}$ that completely suppress the
superfluid density and amplitude of the SC order parameter
correspondingly. $H^{*}_{c2}$ is thus determined self-consistently
and agrees with our initial estimates based on the raw data. We
find that the SC gap still remains open at a finite value at the
effective critical fields $H^{*}_{c2}$. We find that the rate of
gap suppression, $\partial\,{2\Delta(H,T)}/\partial\,H$, is
strongly doping dependent: $-8$ cm$^{-1}$T$^{-1}$ for the
optimally doped crystal rapidly increases with doping to a
surprisingly large $-46$ cm$^{-1}$T$^{-1}$ for over-doped PCCO
with $x = 0.17$.

These observations are in sharp contrast to $p$-doped cuprates
where the field induced suppression of the SC coherence peak
intensity is not accompanied by any observable shift in the gap
magnitude.\cite{BlumbergTl} Moreover, these observations are also
different from the weak shift of the SC mode energy with field in
the Raman spectra on NbSe$_2$, a BCS-type s-wave
superconductor.\cite{Soory1,Soory2} The Doppler shift due to
circulating supercurrents cannot explain the rapid suppression of
the SC gap magnitude. The rate of change of the Zeeman energy with
field is much less than the observed rate of gap magnitude
suppression and can be ruled out as a cause of this suppression. A
possible scenario for the rapid reduction of $H^{2\Delta}_{c2}$
with doping is related to the non-monotonic $d$-wave form of the
SC gap in $n$-doped cuprates for which the points of maximum gap
amplitudes with opposite phase are close to each other in
reciprocal space.\cite{BlumbergNCCO,Takahashi} In this situation,
introduction of vortices leads to QP scattering between points of
large gap amplitude with opposite phase which is self-damaging for
the gap amplitude.

\begin{figure}[t]
\epsfig{figure=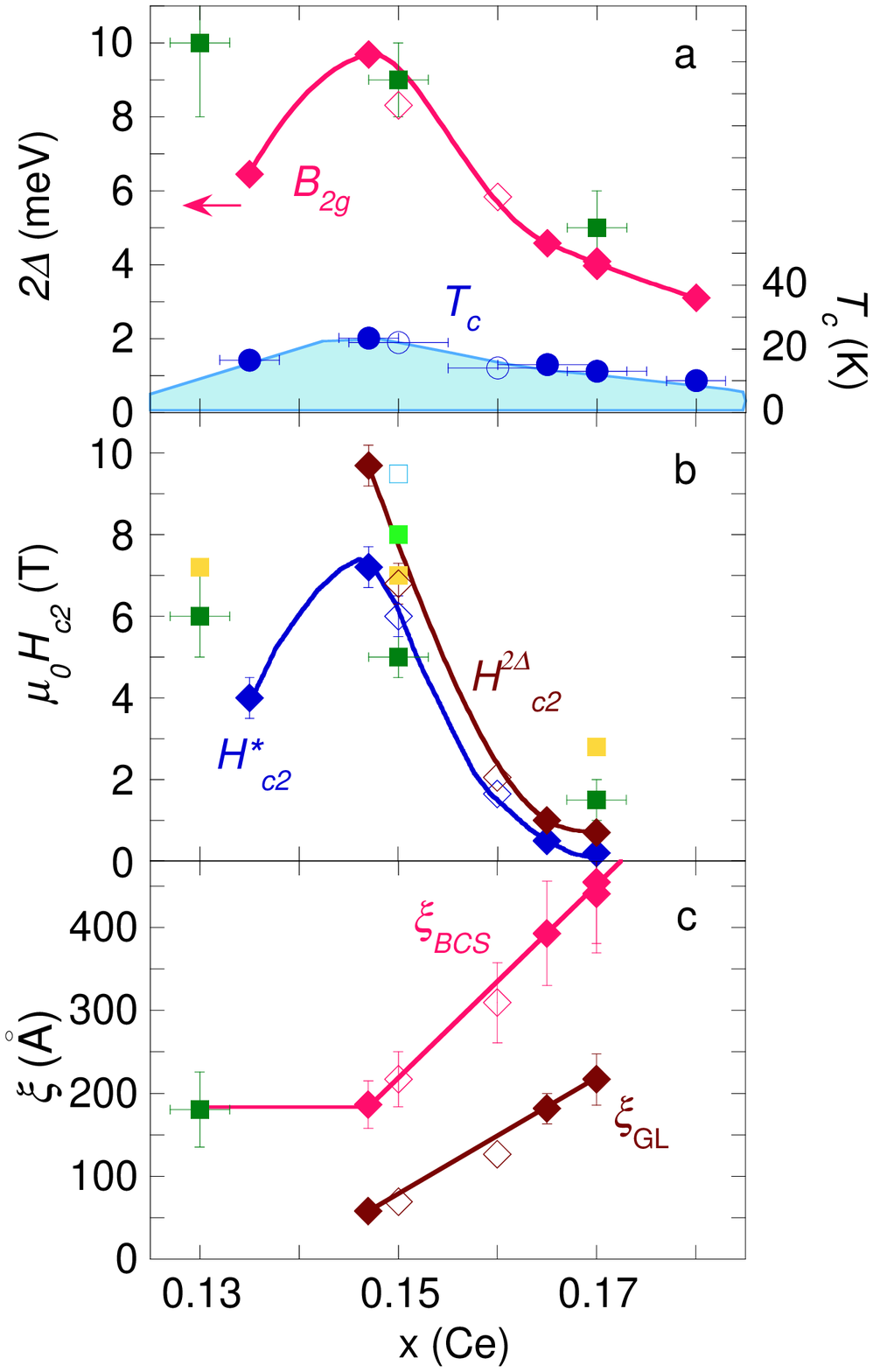,width=75mm}
\caption{(color online) The SC phase diagram of PCCO (filled
diamonds) and NCCO (open diamonds) explored by electronic Raman
scattering in magnetic field. Panels show: (a) T$_c$  (blue
circles), the maximum energy of the $2\Delta$ peak from Raman
spectroscopy, and the distance between coherence peaks from point
contact tunneling spectroscopy (filled dark green
squares);\cite{Amlan,Qazilbash} (b) The doping dependence at 5~K
of the effective upper critical fields $H^{*}_{c2}(x)$ (dark blue)
and the fields suppressing  the gap amplitude
$H^{2\Delta}_{c2}(x)$ (brown) is compared to upper critical fields
obtained from other measurements (squares): tunneling
spectroscopy\cite{amlanprb,Qazilbash} (dark green), Nernst effect
on PCCO films\cite{hamzanernst} (yellow), Nernst effect on NCCO
crystal\cite{Ong} (light blue), and thermal
conductivity\cite{taillefer} (light green); (c) the
Ginzburg-Landau SC coherence length $\xi_{GL}(x)$ (brown) is
compared to the BCS coherence length $\xi_{BCS}(x)$ (red). For the
under-doped sample, the gap value from point contact tunneling is
used to calculate $\xi_{BCS}$ while for all other dopings the gap
values in the Raman $B_{2g}$ channel (Fig. 7a) have been used. The
error bars on the Ce concentrations $x$ are shown only on the
T$_c$ data points to preserve clarity of the figure. All solid
lines are guides to the eye. } \label{Fig.7}
\end{figure}
We turn to Fig.~7 where the upper critical fields, $H^{*}_{c2}(x)$
and $H^{2\Delta}_{c2}(x)$, and the related Ginzburg-Landau SC
coherence length $\xi_{GL}(x) =
\sqrt{\Phi_0/2\pi{H}^{2\Delta}_{c2}(x)}$ ($\Phi_0$ is the fluxoid
quantum) are plotted vs. Ce doping $x$.\cite{underdoped} The
$H^*_{c2}$ value of 7.2 $\pm$ 0.5 T for the optimally doped PCCO
crystal is in agreement with $H_{c2} \approx 8$~T determined by
thermal conductivity\cite{taillefer} and specific heat
measurements.\cite{hamza} This value also agrees with the $H_{c2}$
estimated by tunneling spectroscopy on optimally doped
samples.\cite{Qazilbash,alffprb,alffnature} Nernst effect
measurements estimate $H_{c2} \approx 10$~T for optimally doped
samples,\cite{Ong} consistent with $H^{2\Delta}_{c2}$ = 9.7 $\pm$
0.5 T. The SC phase coherence vanishes at a lower critical field
while the pairing amplitude persists up to a higher field. This
supports our earlier interpretation of the temperature dependence
of the gap in terms of phase fluctuations above $T_c$. With
increasing doping, $H^{2\Delta}_{c2}$ drops by an order of
magnitude reducing to 0.7~T for the $x = 0.17$ over-doped PCCO
crystal.\cite{overdoped} Correspondingly, $\xi_{GL}(x)$ rapidly
increases with doping from 60~\AA \ at optimal doping to 220~\AA \
for $x = 0.17$ PCCO.

We note that the $H_{c2}$ value of optimally doped films from
magneto-resistivity, using the full recovery of resistivity
criterion,\cite{patrick} is consistent with that obtained from
Raman and thermodynamic measurements. However, for the over-doped
film (x $\approx$ 0.17), $H_{c2}$ (at 5 K) based on the appearance
of resistance\cite{patrick} is consistent with $H^{2\Delta}_{c2}$
from the Raman data on similarly doped crystals. This could mean
that vortex flow physics, which plays a dominant role in cuprate
superconductors with short coherence lengths, becomes less
important in the over-doped $n$-type cuprates with longer
coherence lengths.\cite{tinkham,footnote}

In Fig. 7c we show that the values for $H^{2\Delta}_{c2}(x)$ are
related to the SC gap. We compare here $\xi_{GL}(x)$ to the BCS
coherence length $\xi_{BCS}(x) = \hbar v_F/\pi\Delta_{max}(x)$,
where the Fermi velocity $v_F = 4.3 \times 10^5$~m/s is estimated
from ARPES measurements\cite{ARPES} and the lowest temperature SC
gap values $\Delta_{max}(x)$ are used from Raman and tunneling
data.\cite{Qazilbash} The comparison plot reveals that the
$\xi_{BCS}(x)$ trend resembles the doping dependence of
$\xi_{GL}(x)$ and therefore confirms the relation of
$H^{2\Delta}_{c2}(x)$ to the pairing potential $\Delta_{max}(x)$.
It also indicates that $\xi_{GL}(x)$ is still about two times
lower and $H^{2\Delta}_{c2}(x)$ is about four times higher than
expected for a conventional BCS superconductor with corresponding
isotropic gap.

$\xi_{GL}$ for the $n$-doped cuprates is significantly larger than
for their $p$-doped counterparts and that leads to important
differences. First, the size of the Cooper pair is larger than the
average inter-particle spacing: $k_F\xi_{GL}$ ranges between 40
and 150, still smaller than for conventional BCS superconductors
but an order of magnitude larger than for the $p$-doped cuprates.
Second, a larger Cooper pair size requires further pair
interactions to be taken into account and leads to a more
complicated non-monotonic momentum dependence of the SC gap rather
than the simplest $\Delta({\bf{k}}) \propto \cos({k_{x}a}) -
\cos({k_{y}a)}$ $d$-wave form that well describes the gap function
for $p$-doped cuprates with very tight Cooper
pairs.\cite{BlumbergNCCO,Takahashi}


\section{Conclusions}

The superconducting (SC) phase diagram of the electron-doped
cuprates has been explored by Raman spectroscopy. The SC gap
magnitudes in optimally- and over-doped samples are in agreement
with the single particle spectroscopy measurements.
The coupling decreases with increasing Ce concentrations from the
strong-coupling regime for the under-doped sample to a weak-coupling
at optimal doping and beyond.
For the under-doped film, a collective mode in
the Raman data in the SC state implies strong final state
interactions. This collective mode appears in the $B_{2g}$ channel
in contrast to the \emph{p}-doped cuprates where it appears in the
$B_{1g}$ channel. At this stage we can only stress the importance
of this observation that has the potential of giving us an insight
into the pairing mechanism in the SC
state~\cite{Chubukov:99,Chubukov:05}. The full ramifications of
this observation deserve further contemplation and are deferred to
a later publication.

The persistence of SC coherence peaks in the $B_{2g}$ channel for
all dopings implies that superconductivity is mainly governed by
interactions in the vicinity of ($\pm \pi/2a$, $\pm \pi/2a$) regions
of the Brillouin zone (BZ). Moreover, the appearance of SC coherence
as a collective mode in the $B_{2g}$ channel in the under-doped
sample ($x \approx 0.13$) coincides with the appearance of hole-like
carriers near ($\pm \pi/2a$, $\pm \pi/2a$) regions of the BZ as seen
by ARPES in under-doped NCCO.\cite{NCCOARPES} For the sample with Ce
doping $x \approx 0.13$, our $B_{2g}$ Raman data shows that SC
pairing first occurs primarily near the ($\pm \pi/2a$, $\pm \pi/2a$)
regions of the BZ where hole-like carriers reside while the more
numerous electron-like carriers show no comparable sign of SC
pairing in the $B_{1g}$ channel. The presence of hole-like carriers
near ($\pm \pi/2a$, $\pm \pi/2a$) regions of the BZ appears to be
vital for superconductivity in the electron-doped cuprates.

Well-defined SC coherence peaks in the $B_{1g}$ channel occur for
optimally-doped samples and this implies that the electron-like
carriers near the ($\pm \pi/a$, $\pm \pi/4a$) and ($\pm \pi/4a$, $\pm
\pi/a$) regions of the BZ are also gapped at this doping.
Whether the
pairing of electron-like carriers is driven by pairing of
hole-like carriers or occurs independently is an important
question that requires further investigation.

Low energy scattering below the SC coherence peak energies for all
dopings and Raman symmetries is most likely due to nodal QPs and
means that the SC gap has nodes on the Fermi surface. This is
consistent with phase sensitive measurements that find the SC
pairing symmetry to be predominantly
$d_{x^2-y^2}$.\cite{Tsuei,josephjunc} However, the order parameter
is likely to be more complicated than monotonic $d_{x^2-y^2}$
\cite{BlumbergNCCO,Takahashi,luo} given the occurrence of two-band
superconductivity and long SC coherence lengths.

We have also carried out a systematic spectroscopic study of
magnetic field and temperature dependence of the electron-doped
cuprates in the SC state. We have plotted the field and
temperature dependence of the SC gap magnitude and the integrated
intensity of the reduced 2$\Delta$ coherence peaks for various
electron concentrations. From the temperature and doping
dependence of the SC coherence peak, we extract an effective upper
critical field line $H^{*}_{c2}(T, x)$ at which the superfluid
stiffness vanishes. The field dependence of the measured SC gap
reveals an estimate of $H^{2\Delta}_{c2}(T, x)$, an upper critical
field at which the SC amplitude is completely suppressed by field.
For optimally-doped samples, the field effectively suppresses the
superfluid stiffness while the SC amplitude survives higher fields
suggesting a phase fluctuation regime for these samples. We find
that magnetic field applied parallel to the $c$-axis linearly
suppresses the SC gap magnitude at a rapid rate, a phenomenon
different from observations in other type II superconductors in
the clean limit.\cite{Soory1,Soory2,BlumbergTl} This implies a
novel pair-breaking mechanism for $n$-doped cuprates in a magnetic
field.

We find that the SC coherence length, $\xi_{SC}$, increases from
60~\AA \ for optimal doping to 220~\AA \ for the over-doped sample
with $T_c = 13$~K. There appears to be a doping-dependent
crossover in physical properties associated with
superconductivity: relatively robust SC pairing at optimal doping
becomes tenuous in the over-doped regime where field suppresses
the pairing potential at an anomalously large rate while $T_c$
still remains relatively high. We also find that carriers doped
beyond optimal doping remain mainly incoherent and do not
contribute to the Drude conductivity and superfluid density. This
possibly explains the fragility of superconductivity in the
over-doped regime of the electron-doped cuprates.

\section*{Acknowledgements}

The authors thank B. Liang, Y. Dagan, V. N. Kulkarni, Z. Y. Li, C.
P. Hill and M. Barr for assistance with preparation and
characterization of samples. MMQ and RLG acknowledge support of
NSF grants DMR 01-02350 and DMR 03-52735. CAK acknowledges support
from ONR/NRL.

\begin{figure}[t]
\epsfig{figure=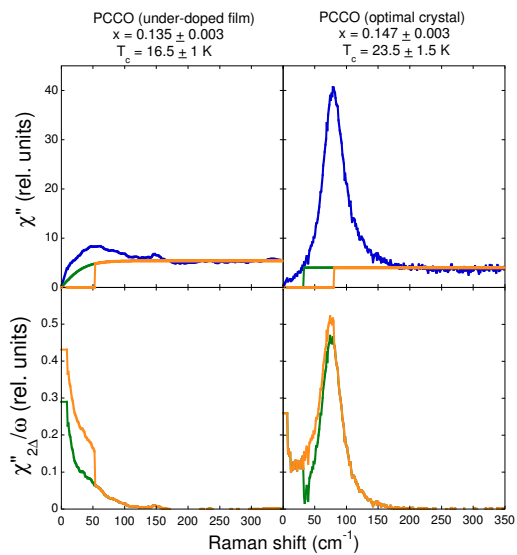,width=68mm}
\caption{(color online) The panels show the main steps in
computing $I_{SC}^{B_{2g}}(x)$ for under-doped PCCO (left column)
and optimally doped PCCO (right column). The details about the
different curves are given in the text. } \label{Fig.8}
\end{figure}

\section*{APPENDIX}

The panels in Fig. 8. show the main steps in the procedure for
computing $I_{SC}^{B_{2g}}(x)$ for two doping concentrations of
PCCO, under-doped PCCO in the left column and optimally-doped PCCO
in the right column. The measured Raman response in the SC state
at lowest temperature is depicted by blue curves (at the lowest
frequency, below 6 cm$^{-1}$, the data is linearly extrapolated to
$\omega$ $\rightarrow$ 0 limit). The yellow and green curves in
the upper two panels are the MFL continua based on two assumptions
respectively: (i) that the continuum vanishes just below
2$\Delta$, and (ii) that it falls off at $\omega_c$ in the same
way as it was derived from the fit to the data just above $T_c$.
In the case of the optimally-doped crystal, the scattering in the SC
state falls off more rapidly at low frequencies than the continuum
based on assumption (ii), and therefore the continuum is truncated
at 30 cm$^{-1}$. The two lower panels show
$\chi''^{B_{2g}}_{2\Delta}/{\omega}$ after the continua are
subtracted in two ways: yellow, using assumption (i) and green,
for assumption (ii). The average of the areas under the two curves
for each sample gives the value of the integrated intensity in the
SC state and one-half the difference between the two areas was
used as an estimate of the uncertainty. The same analysis was
applied to the spectra from all other samples.


\begin{references}
\bibitem{films}
The featureless continuum is stronger for the film samples
compared to the single crystals. This is an indication of
additional disorder and/or possible strain fields.
\bibitem{resonance}
Comparison of $I_{SC}^{B_{2g}}(x)$ with doping-dependence of
1/$\lambda^2$ is relevant if one assumes that the resonant
properties of the Raman vertex, i.e. the optical conductivity near
the laser excitation of 647 nm (1.9 eV), do not change with
doping. Indeed, there is only a weak change in optical
conductivity above 0.8 eV with doping in the relevant doping range
(0.13 $<$ x $<$ 0.18).\cite{arima,onose}
\bibitem{d_vs_s} A fit to the BCS theory does not necessarily mean
that the gap in the over-doped samples has $s$-wave symmetry.
\bibitem{underdoped} Weak SC coherence peak intensity from the
under-doped PCCO film makes it difficult to determine the gap
values for higher fields. At present, we can only estimate the
upper critical field $H^{*}_{c2}$ at which the SC coherence peak
vanishes.
\bibitem{overdoped}
Point contact tunneling\cite{Qazilbash} gives a higher $H_{c2}$
of 1.5~T for over-doped samples because the tunneling measurements
were performed at lower temperature (1.4~K).
\bibitem{footnote} The higher $H_{c2}$ estimates for over-doped
samples extracted from resistivity and Nernst measurements may be
due to $T_c$ distribution in those samples.\cite{patrick,hamzanernst}
\end{references}
\end{document}